\documentclass[
 amsmath,amssymb,
 aps, twocolumn,
 superscriptaddress] {revtex4-1}

\usepackage     {graphicx}      
\usepackage     {color}         
\usepackage		{upgreek}       
\usepackage     {hyperref}      
\usepackage     {braket}        

\definecolor{green}{rgb}{0.45, 0.45, 0}
\definecolor{darkred}{rgb}{0.5, 0, 0}
\definecolor{darkblue}{rgb}{0.1, 0.1, 0.5}
\definecolor{gray}{gray}{0.4}
\hypersetup{
    colorlinks,
    linkcolor=gray,
    citecolor=darkred,
    urlcolor=green,
    pdftitle={Synchronisation in a driven-dissipative hot Rydberg vapor},
    pdfauthor={K. Wadenpfuhl and C. S. Adams},
}

\begin{document}


\title{Emergence of synchronisation in a driven-dissipative hot Rydberg vapor}

\author{Karen Wadenpfuhl}
\email{wadenpfuhl@physi.uni-heidelberg.de}
\affiliation{Joint Quantum Centre (JQC) Durham-Newcastle, Department of Physics, Durham University, DH1 3LE, United Kingdom}
\affiliation{Physikalisches Institut, Universit\"at Heidelberg, Im Neuenheimer Feld 226, 69120 Heidelberg, Germany}

\author{C. Stuart Adams}
\email{c.s.adams@durham.ac.uk}
\affiliation{Joint Quantum Centre (JQC) Durham-Newcastle, Department of Physics, Durham University, DH1 3LE, United Kingdom}

\date{\today}

\begin{abstract}
We observe synchronisation in a thermal (35-60 $^\circ$C) atomic (Rb) ensemble driven to a highly-excited Rydberg state (principle quantum number $n$ ranging from 43 to 79). Synchronisation in this system is unexpected due to the atomic motion, however, we show theoretically that sufficiently strong interactions via a global Rydberg density mean field causes frequency and phase entrainment. The emergent oscillations in the vapor's bulk quantities are detected in the transmission of the probe laser for a two-photon excitation scheme.
\end{abstract}

\maketitle


Nonlinear systems are abundant in nature, where the nonlinearities introduce a range of rich and varied phenomena. Well known is the ability of nonlinear systems to generate multiple steady states, so that the system's state is determined by its past trajectory and hysteresis loops may form. Such multistable states have been observed numerously in biological \cite{Ozbudak2004, Angeli2004, Harrington2013, Newman2010}, mechanical \cite{Chan2001, Badzey2004, Guerra2010}, and atomic systems \cite{Gibbs1976, Hehlen1994, Carr2013, Wade2018}. Nonlinear dynamics and bifurcation theory provide a modelling framework of these phenomena, enabling a fundamental understanding of the underlying processes from within a generalised mathematical framework.

When adding dissipation to a conservative nonlinear system, the resulting dynamics get even richer and the system can support rather unexpected types of stable solutions. Under certain conditions, dissipative systems with nonlinearities can support chaotic behavior \cite{Lorenz1963, Marek1991} or limit cycles and time-periodic solutions \cite{Ruelle2014, Marsden1976}. A Hopf bifurcation may cause the appearance of attractive limit cycles, which leads to self-sustained oscillations of the system. This oscillatory behavior is not imprinted by an external drive but arises fundamentally from the system's dynamics. Such self-oscillating systems have been found to model biological processes \cite{FitzHugh1961, Hodgkin1952, Zhang2012, Colijn2007, Pankavich2020} and physical systems \cite{Lee2011, Wiesenfeld1996, Dreon2022, vanDerPol1926}.

A very curious question regards the behavior of an ensemble of self-sustained oscillators experiencing a form of coupling to another, or to an external force. First studied by Kuramoto for an ensemble of globally coupled oscillators with different natural frequencies \cite{Kuramoto}, it has been found that - under certain conditions - all or a subset of the oscillators begin to lock in frequency and phase \cite{Pikovsky2001, Strogatz2000, Acebron2005}. As a result, a transition towards a sychronised state occurs in the ensemble. This synchronisation transition has been used to explain e.g. the strong lateral vibrations of the Millennium bridge, London, on its opening day \cite{Dallard2001}, though this is contested \cite{Belykh2021}, or the Belousov-Zhabotinsky and other chemical reactions \cite{Kuramoto1984, Ertl1991}. In nature, synchronisation occurs in ensembles of fireflies flashing in unison \cite{Buck1988}, the chirps of snowy tree crickets \cite{Walker1969}, and occasionally in the applause of audiences \cite{Neda2000}.

To further study the emergence of synchronisation and the resulting non-equilibrium dynamics, a simple and easily controllable system with a macroscopic number of coupled oscillators and tunable properties is highly desirable. In the following, we demonstrate that the occurrence of a synchronised phase is expected in a continuously driven, dissipative three-level system with a power law coupling to a mean field, and report on the observation of synchronisation in a hot Rydberg vapor. A surprising, but expected, feature of this system is that oscillations of the bulk quantities remain observable even though the individual constituents are undergoing random motion.


\begin{figure}
    \includegraphics[width=\linewidth]{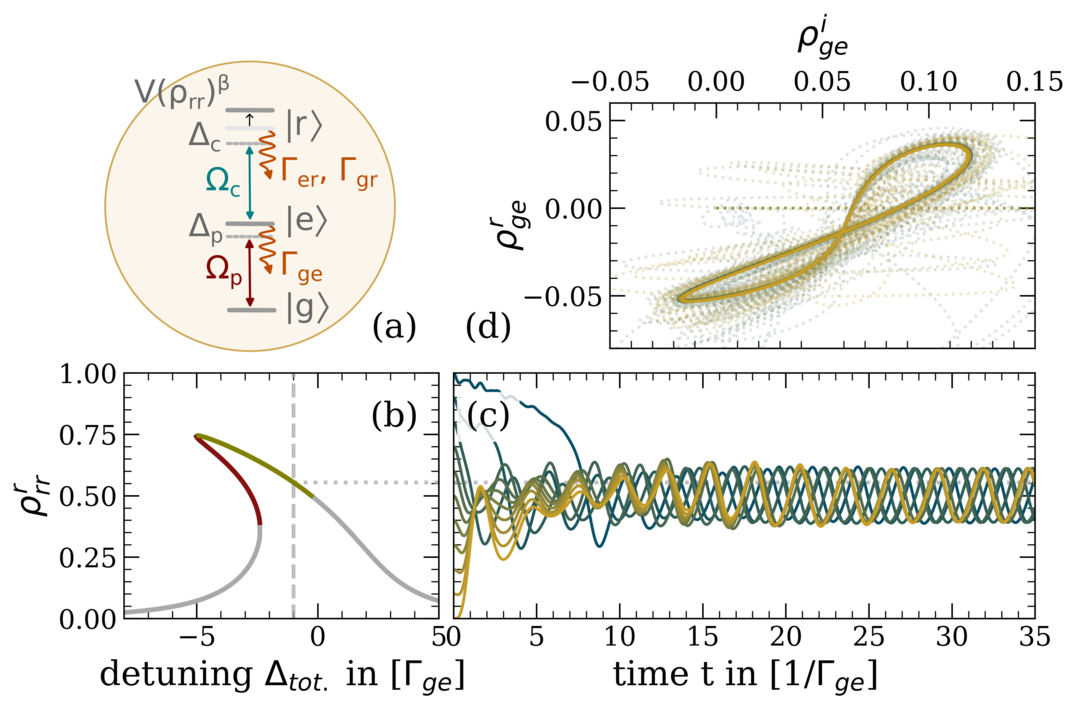}
    \caption{\textbf{Single velocity class model.} The basic model with the relevant parameters is shown in (a). An example steady-state solution of the resulting nonlinear OBEs is shown in (b) where the dark-red steady-state branch is repulsive and green indicates the limit cycle region. For a fixed detuning $\Delta_c/\Gamma_{ge}=-1$, indicated by the dashed line, the time evolution from an initial state $\ket{\Psi}_{t=0}=(1-x)\ket{g}+x\ket{r}$ with $x\in~[0,1]$ towards a limit cycle is shown in (c). For the same time traces, a phase space projection of the limit cycle in the $\rho_{ge}$-plane is shown in (d). The other model parameters were set to: $\Delta_p = 0$, $\Omega_p / \Gamma_{ge}=3.8$, $\Omega_c / \Gamma_{ge}=2$, $V/\Gamma_{ge}=-12$, $\Gamma_{er}/\Gamma_{ge}=10^{-5}$, $\Gamma_{gr}/\Gamma_{ge}=10^{-2}$ and $\beta = 3$.}
    \label{fig:singleVelocityClass}
\end{figure}

Rydberg atoms are known to interact strongly with a power law scaling in distance. This translates into a mean-field approach \cite{deMelo2016} with power law scaling $\beta$ of the Rydberg level shift in Rydberg density $\rho_{rr}$. A similar power law scaling also can be used to model the level shift induced by ionization \cite{Weller2019} or other mean-field inducing mechanisms. Adopting this mean field approach, the resulting equations of motion (EOMs) are formulated for a three-level basis set with coherent driving by $\Omega_x$ and dissipation $\Gamma_{yz}$, see figure \ref{fig:singleVelocityClass} (a). For $\beta\neq0$, the EOMs are nonlinear and their steady states are defined by the roots of a polynomial of order $max(4\beta+1, 1)$ in $\rho_{er}^i$.

The resulting steady-state solutions of the nonlinear EOMs reveal regions of multistability where an odd number of equilibria exist for one set of parameters $\Omega_x, \Delta_x, \Gamma_{yz}, V, \beta$. To extract the stability of the solutions, the spectrum of eigenvalues $\lambda_j$ of the linearisation (Jacobi) is evaluated at the steady state \cite{Hartman2002}. Stability is guaranteed if $\emph{Re}(\lambda_j) < 0$ for the eight non-constant eigenvalues. Consequently, the repulsive branch marked in red in figure \ref{fig:singleVelocityClass} (b) is detected by spectral analysis. However, the steady states indicated in green are also unstable. Here, a Hopf bifurcation occurs where a complex-conjugate pair of eigenvalues $\lambda_j$ crosses the imaginary axis and renders the steady state unstable. As a result, the system is attracted towards a limit cycle which leads to robust self-sustained oscillations of the system parameters in time. Figure \ref{fig:singleVelocityClass} (c) and (d) show that the system is attracted to the same limit cycle for different initial states, but each initial state leads to a different phase in the limit cycle at any fixed time $t$. This freedom of phase in the limit cycle is indicative of a self-oscillating system and fundamentally distinguishes it from a periodically driven system where the phase in the limit cycle is locked to that of the drive. The freedom of phase in the resulting limit cycle has also been described using the language of continuous time crystals \cite{Kongkhambut2022, Liu2023}. The time-crystal interpretation in the context of our experiment is discussed in appendix E of the supplementary material \cite{supplementaryRefs}.

\nocite{Parks1962}
\nocite{Press2007}
\nocite{Sibalic2017}

\begin{figure}
    \includegraphics[width=\linewidth]{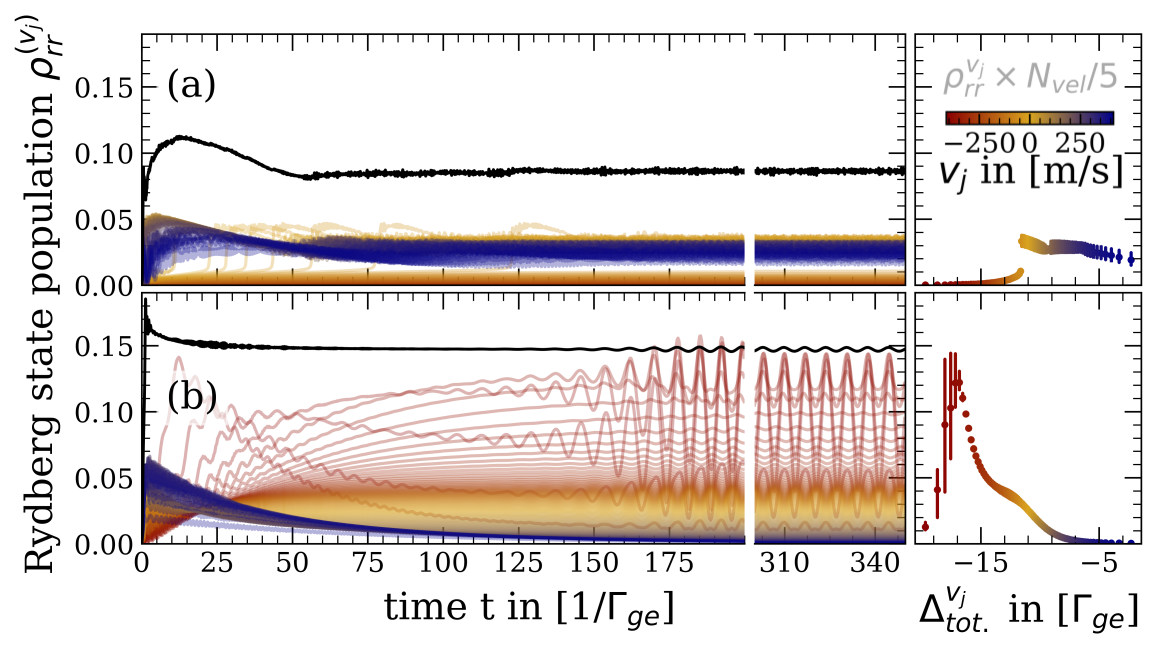}
    \caption{\textbf{Thermal vapor simulation showing emergence of synchronisation.} A thermal vapor simulation for uncoupled (a) and coupled (b) velocity classes shows the emergence of synchronisation via the Rydberg density induced mean field. The time-evolution and corresponding steady-state spectrum are shown on the left and right, respectively. Simulation parameters were $\Omega_p/\Gamma_{ge}=6$, $\Omega_c/\Gamma_{ge}=4$, $\Delta_p=0$, $\Delta_c/\Gamma_{ge}=-11$, $\Gamma_{er}/\Gamma_{ge}=10^{-5}$, $\Gamma_{gr}/\Gamma_{ge}=10^{-2}$, $V/\Gamma_{ge}=-800$, $\beta=2$ and $N_{vel}=101$ velocity classes with equal populations. The atomic velocity distribution corresponds to that of a rubidium vapor on the D$_2$ line at 48 °C.}
    \label{fig:thermalVapor}
\end{figure}

Although optical bistability has been found experimentally in driven-dissipative hot Rydberg vapors \cite{Carr2013}, one would intuitively expect any oscillations in this system to average out due to atomic motion. The motion-induced dephasing for different atomic velocities results in a spread of the natural frequencies of the limit cycles and the phases therein. Although about half of the velocity classes are attracted towards a limit cycle, no macroscopic oscillations can be seen, as shown by the black line in figure \ref{fig:thermalVapor} (a).

However, above argumentation does not account for the spatial dimension of the situation. The Rydberg level shift of any atom in the vapor depends on the spatial Rydberg density of its local environment so that the different velocity classes do not evolve independent of another. Rydberg atoms of one velocity class experience a level shift depending on the Rydberg population of the other velocity classes in the vapor and, in turn, influence the dynamics of these other velocity classes. When taking this global coupling between the velocity classes into account, the resulting dynamics of the vapor is very different as shown in \ref{fig:thermalVapor} (b) [see also App. C in Supp. Mat]. After an initial transient phase, synchronisation sets in where the velocity classes begin to oscillate in lockstep with a single frequency and fixed phase relation. This is possible because the phase of a velocity class within its limit cycle is free and therefore easily adjusted by the mean field. With a growing number of velocity classes oscillating in phase lock, the mean field strength increases which forces even more velocity classes to align their oscillations until eventually a partially or completely synchronised state is reached.

This transition towards a synchronised state of globally coupled oscillators is known since Christiaan Huygens' time \cite{Willms2017} and has since been studied extensively from a mathematical perspective. After the initial work by Winfree \cite{Winfree1967} and Kuramoto \cite{Kuramoto}, the study of synchronisation has been extended to more general forms of the global coupling force \cite{Strogatz2000, Acebron2005} and other situations \cite{Lee2013}. Famous examples where synchronisation is experimentally demonstrated for few oscillators is the synchronisation of pendulum clocks \cite{Willms2017} or metronomes \cite{Pantaleone2002} fixed to a common support which provides the coupling. However, large numbers of globally coupled oscillators with widely tunable properties are not so easily available. Therefore, a hot Rydberg vapor with $\sim \mathcal{O}(10^9)$ atoms in the beam volume, and a somewhat lower number of oscillators, provides an ideal testbed for an experimental study of the synchronisation transition for large numbers of constituent oscillators.


\begin{figure*}
    \includegraphics[width=\linewidth]{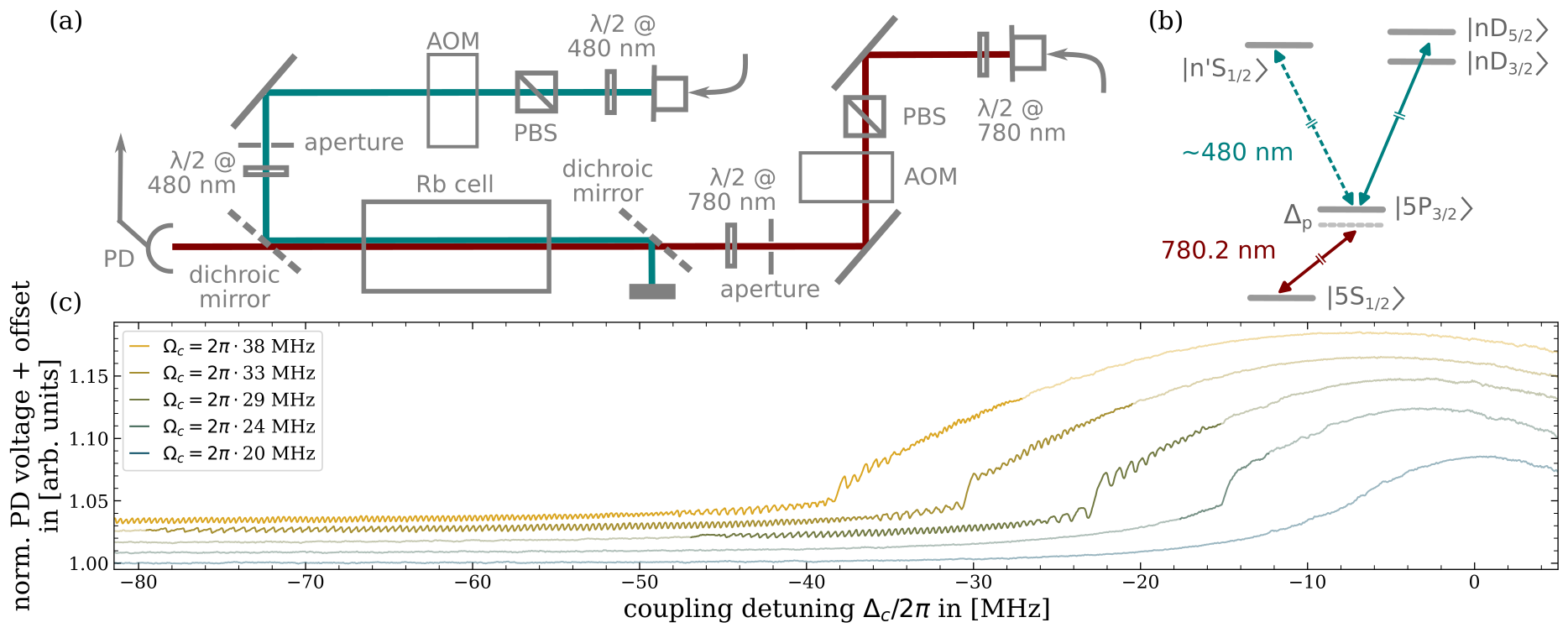}
    \caption{\textbf{Setup and example onset of oscillations.} (a) The counterpropagating probe and coupling lasers are polarisation cleaned with a polarising beamsplitter (PBS) after exiting the fibers. The subsequent acousto-optic modulator (AOM) and aperture are used to remote control the laser powers incident on the heated, 4 cm long rubidium cell. The probe light is detected by a photodetector (PD). (b) shows the relevant level scheme for two-photon spectroscopy in rubidium. The coupling laser addresses either the $\ket{n'S_{1/2}}$ or the $\ket{nD_{m_J}}$ state(s). In (c), an example set of traces obtained for fixed probe Rabi frequency $\Omega_p/2\pi = 191$ MHz and increasing coupling Rabi frequencies is shown. The Rydberg laser is coupled to the $\ket{43D_{5/2}}$ state, and the number density is $\rho_{87Rb} = (4.7 \pm 0.2)\cdot 10^{10}$ cm$^{-3}$. Here, the oscillatory regime is preceded by an onset of bistability.}
    \label{fig:setupAndOscillations}
\end{figure*}

In our experiment, we use $^{87}$Rb number densities of $\rho_{\rm 87Rb} \in [0.1, 6.1] \cdot 10^{11}$ cm$^{-3}$, which corresponds to temperatures from 35 $^\circ$C to 60 $^\circ$C for a vapor of rubidium with natural abundance. The probe laser was locked to a detuning of $\Delta_p / 2\pi = -140$ MHz below the $^{87}$Rb resonance with the intermediate state $\ket{5S_{1/2}, F=2} \rightarrow \ket{5P_{3/2}, F=3}$. The counterpropagating coupling laser was set to scan through two-photon resonance with a $\ket{nS_{1/2}}$ or $\ket{nD_{5/2}}$ Rydberg state at typical scan speeds of up to $2\pi\times 10$ MHz/ms. Typical Rabi frequencies were in range $\Omega_p / 2\pi \in [100, 330]$ MHz and $\Omega_c / 2\pi \leq 35$ MHz for Rydberg states with principal quantum numbers $n$ ranging from 43 to 79. Different beam waists of up to $w \leq 1$~mm and beam waist ratios of $w_p / w_c \approx 2,\ 0.9,\ 0.5$ have been tried, but no direct dependence on the beam waists has been observed. The data presented here was obtained for $w_p = 390\ \upmu$m and $w_c = 440\ \upmu$m. Setup and relevant level scheme are shown in figure \ref{fig:setupAndOscillations} (a) and (b).

Figure \ref{fig:setupAndOscillations} (c) shows a typical series of scans for fixed probe and increasing coupling Rabi frequency. After an onset of bistability in the optical response, a window featuring oscillations in the vapor transmission opens. This synchronisation window widens for a further increase in coupling Rabi frequency. When instead setting the coupling Rabi frequency to a fixed value, the width of the oscillation region decreases with increasing probe Rabi frequency (see also App. D in Supp. Mat.). In the various parameter regimes that were explored experimentally, the synchronisation regime is often preceded by bistability but not necessarily so. We find a strong dependence of the onset of oscillations on the Rydberg state and vapor density. Higher atom number densities require lower Rabi frequencies for the oscillations to set in. This behavior is expected from a synchronisation perspective since larger global coupling strengths require lower mean-field strengths to initiate entrainment.

We observe an onset of synchronisation for coupling to both $n$S and $n$D Rydberg states, though it is easier to explore the behavior and scaling when coupling to D states due to the stronger dipole coupling at similar $n$. The oscillations were also observed when coupling a fourth P or F state with an additional rf field in both the weak and strong driving limit, respectively. In the fully Autler-Townes split regime, oscillations occurred as long as the Rydberg population was high enough. The presence of synchronisation is therefore neither a purely three-level phenomenon, nor does it depend on the orbital angular momentum of the Rydberg state.

\begin{figure*}
    \includegraphics[width=\linewidth]{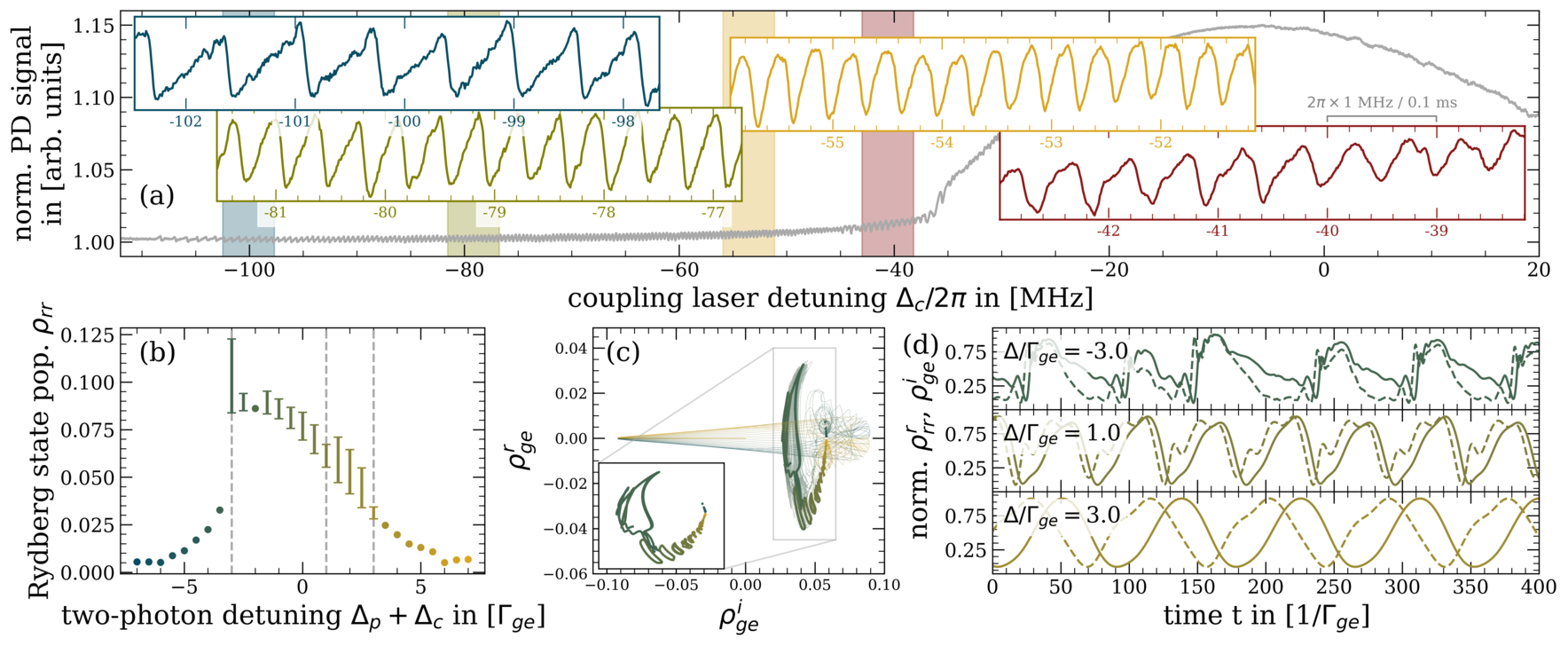}
    \caption{\textbf{Change in oscillation shape and frequency along coupling laser scan.} (a) shows the oscillation region for a scan across resonance with $\ket{43D_{5/2}}$ at $T=(52.0 \pm 0.5)\ ^\circ$C with $\Omega_p / 2\pi = 191$~MHz, $\Omega_c / 2\pi = 37$~MHz, and a scan rate of $2\pi\times 10$ MHz/ms. The colored insets show a zoom-in of the trace in the color-shaded regions, each of width $2\pi \times 4.8$ MHz. Different shapes of the oscillations can be distinguished. (b) Pointwise integrated spectrum with errorbars denoting the amplitude of the oscillations. The time evolution towards a limit cycle is shown in (c) with the inset showing only the limit cycles approached after an integration time of $t = 5000 \Gamma_{ge}^{-1}$. In (d), the oscillations in Rydberg population $\rho_{rr}^r$ (solid) and in the imaginary part of the coherece $\rho_{ge}^i$ (dashed) are shown. The case $\Delta = -3\Gamma_{ge}$ did not approach a limit cycle within the maximum integration time but behaves similar to a system near a strange attractor. The simulation assumes a thermal vapor with $N_{vel}=101$ velocity classes with equal populations, $\Omega_p = 1.5$, $\Omega_c=1$, $\Delta_p=0$, $\Gamma_{er}=10^{-6}$, $\Gamma_{gr}=10^{-3}$ and $V=-300$, in units of $\Gamma_{ge}$, and $\beta=2$.}
    \label{fig:osciShapeComparison}
\end{figure*}

With all system parameters held constant and fixed laser detunings, the synchronised state persist on timescales on the order of minutes and the oscillations maintain their shape. Analysis of a time trace reveals a narrow frequency peak with a spectrum of weaker, higher harmonics (also shown in App. D of Supp. Mat.). The oscillation frequency $\nu_{osc}$ of the first peak was usually observed to lie between 10 kHz and 25 kHz, though persistent oscillations of up to 43 kHz were measured. In Fig. \ref{fig:setupAndOscillations} (c) one can see that the oscillation frequency varies along the coupling laser scan. As a general trend, an increase in oscillation frequency $\nu_{osc}$ with increasing Rabi frequencies was observed. Additionally, the formation of several separate synchronisation regions, typically with a different range $\nu_{osc}$ but similar shapes of the oscillations along the region, has been found. This is also visible in \ref{fig:setupAndOscillations} (c) where the two regions share a boundary at $-\Delta_c / 2\pi \approx 26,\ 36,\ 48$ MHz for $\Omega_c / 2\pi = 29,\ 33,\ 38$ MHz, respectively.

Figure \ref{fig:osciShapeComparison} (a) shows the change in oscillation shape and frequency with increasing $\Delta_c$. Each highlighted segment samples the time dependence at a particular detuning as the laser frequency is scanned in time slowly relative to $\nu_{osc}$. The rightmost zoom-in (red) belongs to the next synchronisation region beginning at $\Delta_c / 2\pi \approx -45$ MHz. It shows again the sawtooth-like shape at its lower frequency end that can also be seen in the two leftmost insets. Figure \ref{fig:osciShapeComparison} (b) - (d) show results obtained with the thermal vapor simulation. The imaginary part of the coherence $\rho_{ge}^i$ shown in (c) and (d, dashed) is linearly proportional to the probe laser transmission via the probe electric susceptibility \cite{Fleischhauer2005}. Two limit cycle regions appear in the spectrum (b), though a cross-section of phase space shows that the case $\Delta_c / \Gamma_{ge} = -3$ is not a limit cycle but resembles a system near a strange attractor. Generally, the thermal vapor model shows regions of multistability which implies that the pointwise integration technique in (b) cannot accurately model a laser scan. This is because the thermal vapor system's trajectory depends on its past state and the attractor it is drawn to, which pointwise integration does not account for.

The thermal vapor model reproduces the observed experimental behavior phenomenologically. This includes changes in the width of the synchronisation region with changes in $\Omega_{\rm p}$ or $\Omega_{\rm c}$ and the earlier onset of oscillations at lower $\Omega_{\rm c}$ for increasing interaction strengths $V$ as shown in the data of Appendix D in the Supp. Mat., as well as the expected shape of the oscillations. Therefore, we attribute the emergence of macroscopic oscillations in the bulk response of a hot Rydberg vapor to a Kuramoto-like synchronisation transition for sufficiently large global coupling strengths. Possible mechanisms causing the power law scaling of the Rydberg density mean field are Rydberg interactions \cite{deMelo2016} or charge-induced Stark shifts due to ionisation \cite{Weller2019}, though other effects could possibly lead to similar power-law scaling behaviors.


In summary, we observe the transition towards synchronisation in a strongly driven, dissipative, hot Rydberg vapor. The observed changes of the synchronised region with variation of the Rabi frequency, vapor density, and interaction strength is reproduced by a theoretical model extended to a thermal vapor simulation. The model's nonlinearity leads to the emergence of attractive limit cycles for individual velocity classes through a Hopf bifurcation. Under the influence of global coupling through the shared Rydberg density, the constituent oscillating velocity classes synchronise in a thermal vapor, which leads to periodic oscillations of the vapor's bulk quantities. The resulting synchronised phase is robust and stable, and therefore ideally suited for an experimental investigation of the emergent non-equilibrium phase of matter. It provides a simple platform for the study of synchronisation in a nonlinear system with a truly macroscopic number of oscillators.

\emph{Author's note:} During completion of this work, two other reports of oscillations in a continuously driven hot Rydberg vapor were reported. In \cite{Ding2023}, the oscillations are of a transient nature and the probe Rabi frequency is significantly lower than in this work. The authors attribute the origin of the limit cycles to spatial inhomogeneities and clustering of Rydberg atoms. In \cite{Wu2023}, the experimental parameter regime is similar to this work. The limit cycles are attributed to a competition for Rydberg population between energetically closely spaced Rydberg states.


\begin{acknowledgments}

K.W. acknowledges insightful discussions with Finn M\"unnich and Matt Jamieson, and thanks Matthias Weidem\"uller. C.S.A. acknowledges fruitful discussions with Dong-Sheng Ding. The authors furthermore thank Lucy Downes, Max Festenstein, Oliver Hughes, and Kevin Weatherill. Financial support was provided by the UKRI, EPSRC grant reference number EP/V030280/1 (“Quantum optics using Rydberg polaritons”).
\end{acknowledgments}


\bibliography{bibliography}


\pagebreak
\ \\
\newpage

\onecolumngrid
\appendix

\section*{Supplementary Material for \\ Emergence of synchronisation in a driven-dissipative hot Rydberg vapor}

In the following supplementary material, we spell out the non-linear equations of motion for the three-level system in \ref{app:threeLevel}. In \ref{app:twoLevel}, we show that no Hopf bifurcation can occur in an effective two-level system after elimination of the intermediate state. The thermal vapor integration scheme is described in \ref{app:thermalVaporIntegrationScheme}, and supplementary experimental results are presented in \ref{app:experimentalSupplement}. Experimental results on the phase of the oscillations and comments on the phase invariance of continuous time crystals are discussed in \ref{app:TimeCrystals}.

\section{3-level model}
\label{app:threeLevel}
The three-level model with a Rydberg density-dependent power law level shift $V \rho_{rr}^n$ leads to the nonlinear equations of motion
\begin{subequations}
\label{eqn:3LevelEOMs}
\begin{align}
\dot{\rho}_{gg} = &\ -\Omega_p Im(\rho_{ge}) + \Gamma_{ge}\rho_{ee} +\Gamma_{gr}\rho_{rr} \\
\dot{\rho}_{ee} = &\ + \Omega_p Im(\rho_{ge}) - \Omega_c Im(\rho_{er}) - \Gamma_{ge}\rho_{ee} + \Gamma_{er}\rho_{rr} \\
\dot{\rho}_{rr} = &\ +\Omega_c Im(\rho_{er}) -(\Gamma_{gr}+\Gamma_{er})\rho_{rr} \\
\dot{\rho}_{ge} = &\ -\frac{i}{2}\Omega_p(\rho_{ee} - \rho_{gg}) +\frac{i}{2}\Omega_c\rho_{gr} -i\Delta_p\rho_{ge} - \frac{\Gamma_{ge}}{2}\rho_{ge} \\
\dot{\rho}_{er} = &\ -\frac{i}{2}\Omega_c(\rho_{rr} - \rho_{ee}) -\frac{i}{2}\Omega_p\rho_{gr} -i\left(\Delta_c-V(\rho_{rr})^\beta\right)\rho_{er} \\ \notag
&\ - \frac{\Gamma_{ge}+\Gamma_{er}+\Gamma_{gr}}{2}\rho_{er} \\
\dot{\rho}_{gr} = &\ -\frac{i}{2}\Omega_p\rho_{er} +\frac{i}{2}\Omega_c\rho_{ge} - \frac{\Gamma_{gr}+\Gamma_{er}}{2}\rho_{gr} -i\left(\Delta_p+\Delta_c -V(\rho_{rr})^\beta\right) \rho_{gr}
\end{align}
\end{subequations}

The steady state solutions are defined via a polynomial of order $max(4\beta+1, 1)$ in $\rho_{er}^i$. The steady state is stable if $\emph{Re}(\lambda_j) < 0$ holds for all eight non-constant eigenvalues of the Jacobi defined by \ref{eqn:3LevelEOMs}. Attractive limit cycles may form at a Hopf bifurcation, i.e. when a complex-conjugate pair $\lambda_k, \bar{\lambda}_k$ crosses the imaginary axis. Such a Hopf bifurcation occurs in the three-level system for certain parameter regimes.

\section{Effective 2-level model}
\label{app:twoLevel}
In an effective two-level system, the intermediate state is eliminated on the grounds of a detuning of the probe laser far off resonance with the intermediate state. The resulting equations of motion
\begin{subequations}
\label{eqn:2LevelEOMs}
\begin{align}
\dot{\rho}_{gg} = &\ -\Omega_p Im(\rho_{gr}) + \Gamma \rho_{rr} \\
\dot{\rho}_{rr} = &\ +\Omega_p Im(\rho_{gr}) - \Gamma \rho_{rr} \\
\dot{\rho}_{gr} = &\ -\frac{i}{2}\Omega (\rho_{rr}-\rho_{gg}) -\frac{1}{2}\Gamma \rho_{gr} - i\left(\Delta - V \rho_{rr}^\beta\right)\rho_{gr}
\end{align}
\end{subequations}

can be solved for the steady state by setting the left hand side to zero. The steady state solutions are defined via the roots of the polynomial
\begin{equation}
\label{eqn:polyForBoundednessOfRoots}
0 = -\frac{\Omega}{2} + \rho_{gr}^i\cdot \frac{2}{\Gamma} \left[ \frac{\Gamma^2}{4} + \frac{\Omega^2}{2} +\left( \Delta - V_l \left( \frac{\Omega}{\Gamma} \rho_{gr}^i \right)^\beta \right)^2 \right]
\end{equation}

The characteristic polynomial of the Jacobi $J$ of the system defined by equations \ref{eqn:2LevelEOMs} has three constant eigenvalues $\{0, 0, -\Gamma\}$ and a cubic term
\begin{equation}
\label{eqn:JacobiCubic}
\tilde{\chi}[J] ( \lambda ) =  \left[\begin{array}{l}
\lambda ^3 + 2 \lambda ^2 \Gamma\\

+\lambda\left(\Omega^2+\Delta^2+\frac{5}{4}\Gamma^2-2 (\beta+1) V_l \Delta \left(\frac{\Omega \rho_{ge}^i}{\Gamma}\right)^\beta+(2\beta+1) V_l^2 
\left(\frac{\Omega \rho_{ge}^i}{\Gamma}\right)^{2\beta}\right) \\

+\frac{\Gamma}{4} \left(4\Delta^2+2 \Omega^2+\Gamma^2-8 (\beta+1) V_l \Delta \left(\frac{\Omega \rho_{ge}^i}{\Gamma}\right)^\beta+4 (2\beta+1) V_l^2
\left(\frac{\Omega \rho_{ge}^i}{\Gamma}\right)^{2\beta}\right)
\end{array}\right]
\end{equation}

The Routh-Hurwitz criterion \cite{Parks1962} for a cubic $\mathcal{P}(x)~=~x^3+a_2x^2+a_1x+a_0$ states that all eigenvalues of $\mathcal{P}$ have a negative real part except for one purely imaginary pair iff $a_2 > 0$, $a_0 > 0$ and $a_2 a_1 - a_0 = 0$ are true. If these conditions are satisfied, a Hopf bifurcation occurs.

From \ref{eqn:JacobiCubic} one sees immediately that $a_2 = 2\Gamma > 0$ is satisfied for any $\Gamma > 0$, as we assume in a dissipative system with non-vanishing population loss from state $\ket{r}$. Looking at the last of the three conditions, the equality, then one finds that $a_2 a_1 - a_0 = a_0 + \Gamma (\Omega^2 + 2\Gamma^2) = 0$ cannot be satisfied if the condition $a_0 > 0$ is true. Therefore, no complex conjugate pairs of eigenvalues crosses the imaginary axis and no Hopf bifurcation occurs in the effective two-level model.

\section{Thermal vapor integration scheme}
\label{app:thermalVaporIntegrationScheme}
In a thermal vapor model, the different velocity classes $\textbf{v}_j$ do not evolve independent of another. This is because the Rydberg atom density generates an effective mean field that all velocity classes couple to and are influenced by. Simple integration of the individual velocity classes therefore cannot account for the Rydberg density produced by other velocity classes, and the integration scheme has to be adapted accordingly. The mean field Hamiltonian for a given velocity class $\textbf{v}_j$ changes from $\mathcal{H}^{\textbf{v}_j} \propto (\rho_{rr}^{\textbf{v}_j})^\beta\ket{r_{\textbf{v}_j}}\bra{r_{\textbf{v}_j}}$ to  $\mathcal{H}^{tot}_{\textbf{v}_j} \propto (\rho_{rr}^{tot})^\beta\ket{r_{\textbf{v}_j}}\bra{r_{\textbf{v}_j}}$. To this end, an adapted rk4 integration scheme \cite{Press2007} has been implemented. After preforming a single rk4 time step for all velocity classes, the Rydberg population of the vapor is computed as the weighted sum over the Rydberg population of all velocity classes. Then, the density-dependent level shift $V(\rho_{rr}^{tot})^\beta$ is adjusted in the EOMs for all velocity classes and the next integration step is performed. A matrix-based implementation of the integration scheme in Python allows for a simple parallel computation of one time step for all velocity classes at once.

\section{Supplementary experimental data}
\label{app:experimentalSupplement}

\begin{figure}[b]
    \includegraphics[width=\linewidth]{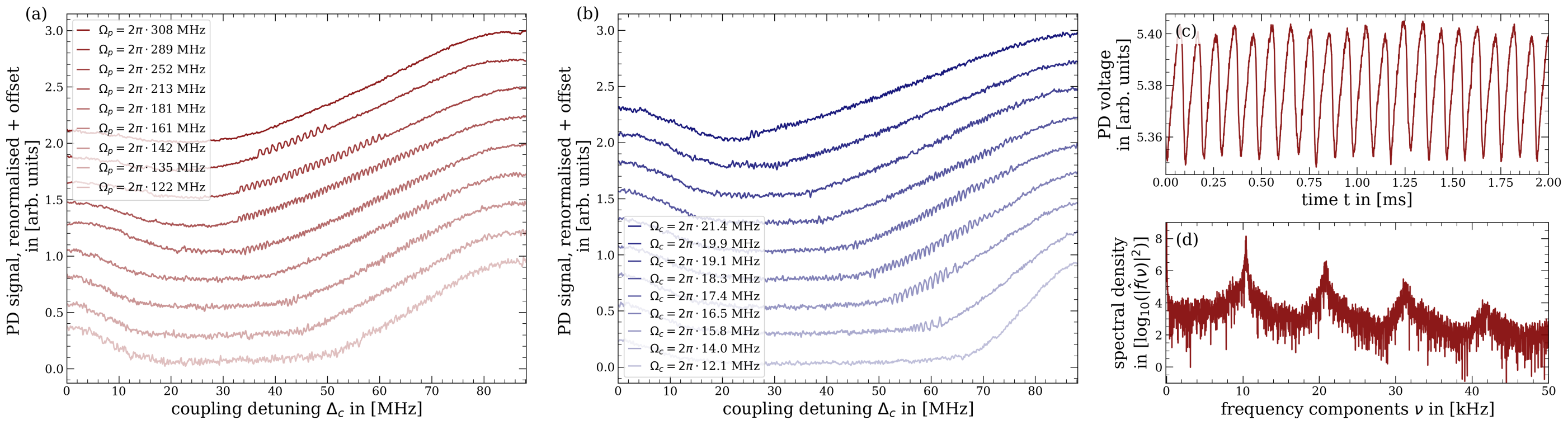}
    \caption{\textbf{Change in width of synchronisation window and spectral analysis of oscillations.} (a) and (b) show example traces of the synchronisation region with coupling to the $\ket{50D_{3/2}}$ state at $\Delta_c / 2\pi \approx 0$ MHz and the $\ket{50D_{5/2}}$ state at $\Delta_c / 2\pi \approx 93$ MHz. (a) For the coupling Rabi fequency held constant at $\Omega_c = 2\pi \times 18$ MHz, the width of the synchronisation window decreases with increasing probe Rabi frequency and the oscillation frequency reduces. When instead keeping the probe Rabi frequency $\Omega_p = 2\pi \times 160$ MHz constant (b), the width of the synchronisation region increases and so does the oscillation frequency. In these examples, the synchronisation region is not preceded by or overlaid with an optical bistability. When keeping all system parameters constant, the oscillations persist and maintain their shape as shown in (c). The resulting spectral intensity (d), plotted in log space, shows a pronounced frequency peak at $\nu_{osc} = 10.4$ kHz and weaker higher harmonics.}
    \label{fig:regionScalingAndSpectralAnalysis}
\end{figure}

Figure \ref{fig:regionScalingAndSpectralAnalysis} shows the observed behavior of the synchronisation window as well as the change in oscillation frequency when keeping coupling (a) or probe (b) Rabi frequency constant and changing the respective other. The observed closing and opening of the synchronisation window, as well as the change in oscillation frequency, are reproduced by the thermal vapor model. The scaling behavior in the thermal vapor model is indirectly inherited from the three-level single velocity class model, which initially seeds the oscillations.

An example time trace for all system parameters held constant is shown in (c) and the corresponding spectral density in (d). This Fourier transform analysis was used to determine the oscillation frequency $\nu_{osc}$.

\begin{figure}
    \includegraphics[width=\linewidth]{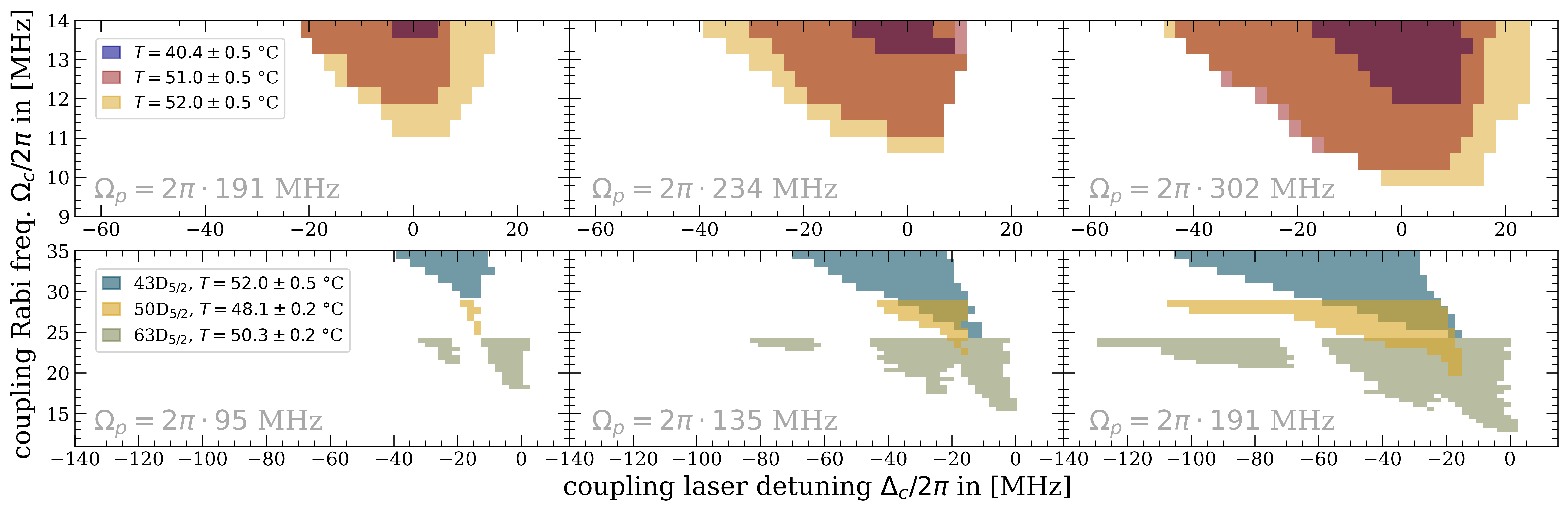}
    \caption{\textbf{Dependence of synchronisation region on vapor density and Rydberg state.} The upper row shows the width of the synchronisation region for the $\ket{79D_{5/2}}$ state at three different temperatures and therefore vapor densities. The other parameters were held constant across the measurements. The width of the region leading to synchronisation, and the minimum coupling Rabi frequency $\Omega_c$ depend on the vapor density $\rho$ and probe Rabi frequency $\Omega_p$. The lower row shows the oscillation region for three different Rydberg states at similar vapor densities.}
    \label{fig:synchronisationRegionScaling}
\end{figure}

The upper row of figure \ref{fig:synchronisationRegionScaling} shows the synchronisation window for three different vapor densities but otherwise identical system parameters with a coupling to the $\ket{79D}$ states. The vapor temperatures were varied from $T_1 = (40.5 \pm 0.5)\ ^\circ$C (blue) to $T_2 = (51.0 \pm 0.5)\ ^\circ$C (yellow) and $T_3 = (52.0 \pm 0.5)\ ^\circ$C (red). By varying the vapor temperature, the critical coupling Rabi frequency $\Omega_c$ where synchronisation sets in is shifted for fixed probe Rabi frequencies and Rydberg state. Therefore, a direct dependence of the coupling strength $V$ on the vapor density can be inferred. A higher vapor density leads to stronger effective Rydberg interactions and ion densities at the same Rabi frequencies due to the reduced interatomic spacing. For higher vapor densities, the onset of synchronisation is therefore shifted to lower Rabi frequencies where the critical Rydberg atom density is then located.

In the lower row of figure \ref{fig:synchronisationRegionScaling}, the synchronisation window is shown for different combinations of Rydberg state but similar vapor densities. At the various probe Rabi frequencies $\Omega_p$, the minimum coupling Rabi frequency required for an onset of synchronisation is different for all three Rydberg states. Higher principal quantum numbers $n$ need lower Rabi frequencies for meeting the synchronisation threshold. The states were selected by Rydberg interaction strength since the $\ket{43D_{5/2}}$ state is on a F\"orster resonance, leading to higher Rydberg interactions than for the $\ket{50D_{5/2}}$ state \cite{Sibalic2017}, while the ionisation cross-section of the lower-$n$ Rydberg state is lower \cite{Weller2019}. The $\ket{63D_{5/2}}$ state has the highest interaction strength and collisional ionisation rates. A direct effect of the F\"orster resonance of the $\ket{43D_{5/2}}$ state cannot be seen in the data.

In the green $\ket{63D_{5/2}}$ data, the occurrence of spectrally separate synchronisation regions is clearly visible.

\vspace{.2cm}

\section{Phase invariance and continuous time crystals}
\label{app:TimeCrystals}

\begin{figure}
    \includegraphics[width=\linewidth]{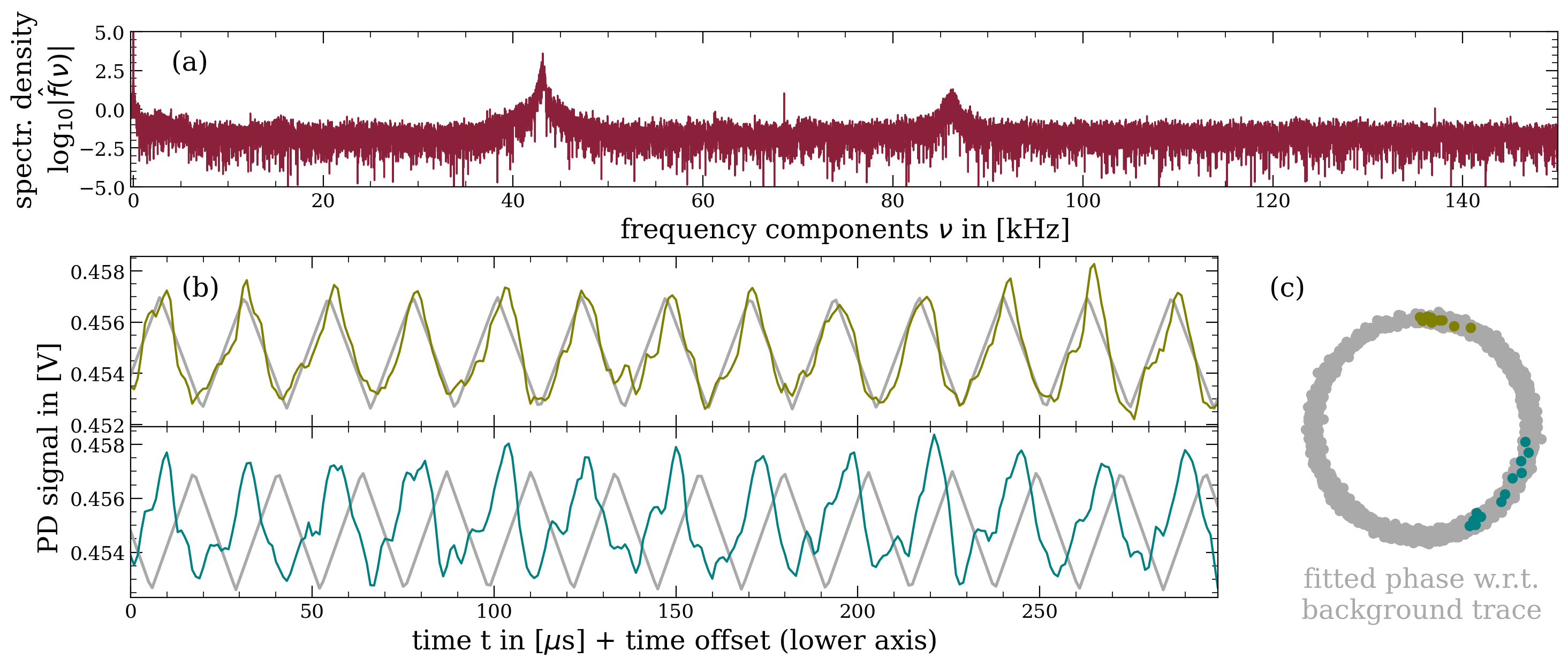}
    \caption{\textbf{Phase drift in the limit cycle during time sequence.} (a) The frequency spectrum of oscillations on the $\ket{63D_{5/2}}$ state with an oscillation frequency $\nu_{osc}$ of 43.1 kHz. Two segments of the 0.1 s long data trace are shown in (b) together with a triangular reference waveform of frequency $\nu_{osc}$ in gray. The bottom time segment shows a phase shift relative to the reference waveform and a change in oscillation frequency along the length of the segment. This leads to a drift of the phase of the signal relative to the reference, also shown in (c), where the extracted phases of the entire sequence are shown in gray. The radius is proportional to the measured oscillation frequency.}
    \label{fig:phaseDriftInLimitCycle}
\end{figure}

The oscillatory phase can be understood as a continuous driven-dissipative time crystal. However, it is experimentally challenging to establish the required phase invariance of the limit cycle within a single realisation. Figure \ref{fig:phaseDriftInLimitCycle} shows a drift in the limit-cycle phase with respect to a reference waveform over time. The observed drift could correlate with uncertainties in the experimental parameters. Experimental uncertainties therefore make it difficult to attribute observed changes in the phase for different realisations solely to the freedom of phase of a time crystal.

\end{document}